\documentstyle[12pt]{article}
 
\begin{document}
\begin{center}
{\Large {\em And the Winner Is . . .} \\
Predicting the Outcome of the 150m Showdown} \\
\vspace{3mm}
J.\ R.\ Mureika\\
{\it Department of Computer Science \\
University of Southern California \\
Los Angeles, CA~~USA~~90089}\\
\end{center}

In a mere 9.84 seconds on July 27, 1996, Donovan Bailey performed multiple 
tasks at once:
he won Olympic Gold, earned the title of ``World's Fastest Man'', 
defeated one of the best 100m final fields ever in
record--breaking time, restored Canadian sprinting to world dominance,
forever banished the tainted poltergeist of Ben Johnson, and shut
out the United States from the medal positions.  Meanwhile, in 19.32 
seconds, Michael Johnson performed a similar number of tasks: he smashed
beyond all recognition the 200m world record in a time faster than anyone
though possible, he recorded blazing splits (10.12s, 9.20s), and (albeit
due in part to the poor math skills of some sports commentators) 
his time evoked the thought in many minds that {\it he} was now the rightful 
heir to the title recently bestowed on Bailey.

Thanks to this hooplah, the Ottawa--based Magellan Group 
has set the stage for a match between these two one--man sprint powerhouses.
On Sunday, June 1st at Skydome, almost 10 months after their respective
world record performances, Donovan Bailey and Michael Johnson will speed
over 150m in what is perhaps to be the most hyped Canada {\it vs} US sprint 
showdown since the 100m final in Seoul on September 24, 1988.  The question
on most minds is naturally: {\it who will win?}

There are numerous ways to guess the end result.  One can consider the
training of each individual.  Bailey is a 100m specialist; does he have
the power to generate world--class speed around a turn, then sustain it
over a straight equal in length to his event?  His best marks at 200m 
stand as 20.76s (20.39s wind--assisted) in 1994.  Johnson, on the other hand,
is a 200/400 doubler.  Does he have the sheer power to match Bailey over
half the race?  His 100m PB is 10.09s, also from 1994.  
Surely, if nothing else, this problem almost borders
on the philosophical!  It has ringings of: ``What happens when an 
irresistible force meets an immutable object?''

One can also look at the numbers that came out of Atlanta.  Bailey was the
last out of the blocks (reaction time of $+0.174$s), reached
an astounding maximum velocity of 12.1 m/s at 60m, and accounting for
tail--wind (+0.7 m/s), ran faster than did Ben Johnson at the Rome WC 
of 1987.  Michael Johnson recorded the previously mentioned splits of 
10.12s/9.20s, led the field at the line by over 0.3s in a {\it sprint}, 
and not counting his earlier 19.66s from the US Trials, broke the 
long--standing record of Pietro Mennea by 0.4s.  

Statistician Robert Tibshirani of the University of Toronto performed such
an analysis, taking into account a 1973 model of Mathematician
Joseph Keller (who in turn based his model on the 1920s work of 
Physiologist A.\ Hill).  It was found that, with a modification
to Keller's model, Bailey would defeat Johnson by a margin of 0.03s--0.22s
in 95 out of 100 races, or at a 95\% confidence level 
(statistically--speaking).

However, Tibshirani noted tha his  model did not take into account the
energy lost by a sprinter (in this case, Bailey) as he rounds the curve.
The Skydome track will have a 50m curve + 100m straight configuration,
so such a model is limited by this non--linearity.  What to do?

Contact a Physicist who has a love of track and field!  Falling into such
a category, I have tried my hand at prediction using a crystal ball of
vector--addition.  Without sounding too much like a 
Classical Mechanics class, let me briefly outline the problem and solution.
The model used
by Tibshirani assumes that the sprinter can exert a time--dependent
force $f(t) = f-ct$; that is, as time progresses, he fatigues.  If we
know certain information about the race ({\it e.g.} splits, speeds),
we can determine the values of $f$ and $c$.  However, the sprinter
also feels internal resistances which sap his strength (anaerobic 
exhaustion, for example), so combining terms for all these effects,
we can obtain a differential equation for his net acceleration.

This works fine for a linear race (100m), but what of the turn?  
A sprinter of mass $m$ rounding a turn of radius $R$ feels a force of 
$mv(t)^2/R$.  Of course, in order to stay on the track (a useful strategy),
he must combat this with his energy reserves.  The 
centrifugal force $mv(t)^2/R$ acts perpendicular to the forward motion
(dependent on the speed of the runner at time $t$),
so we have to add the force terms as vector, the root of the sum of the squares
$f(t)^2$ and $m^2v(t)^4/R^2$.  So far so good, but such an assumption
overcompensates for the curve, that is, it predicts 200m times which
are well above what one would expect for world class performances.
The key is to remember that a sprinter leans into the turn, letting
gravity do some of the work combating the centrifugal force.  So, 
a determination of exactly what fraction $p$ of this force the runner
actually feels can help predict times run on curves.


\section{The 150m Showdown}

The 150m race will be run as a 50m curve + 100m straight.  While the 
exact configuration has yet to be decided, the best choice would be
to have a curve of fairly large radius (larger than for regular indoor
tracks).  From floor plans for the event, it seems as if the curve
will be close in size to those of outdoor tracks.  I'll assume that the
race would be run in the equivalents of lane 3 and 4.  While the
model does not account for wind assistance (Bailey's +0.7 m/s, and
Johnson's +0.4 m/s), it's reasonable to assume that each athlete
is stronger than last year and capable of moving slightly faster.

Using our model, we can take a stab at how Bailey might be able to handle
this race.  The value of $p$ we should use is still up in the air, but it
seems likely that it won't be small.  After all, Bailey is a 100m specialist,
and wouldn't handle turns very well.  So, we'll take $p$ to be either
0.6, 0.7, or at the worst, 0.8 (these are the squares of the percentage
of centrifugal forces felt, which would mean we're roughly considering between
75\% and 90\% of the force).  

 Michael Johnson's 50m split in his Olympic 200m
final was about 6.4s, so assuming similar conditions, Bailey clearly
leads off the curve.  After this point, it's hard to determine how Johnson
would handle the straight.  His split of 10.12s was run on a complete turn,
so we could guess that he could shave off about 0.05s.
This would put his Skydome split at about 10.07s (slightly quicker than
his 1994 PB).  Continuing this logic,
let's guess that he'd be able to hold a greater speed over the straight,
and clock in between 0.05--0.10s faster than his Atlanta 150m split of
14.83s.  Since Johnson is probably a more  consistent  curve runner, we
could assume that his time doesn't vary more than 0.01s from lane to lane.
If this is the case, then Johnson could optimistically clock between
14.73s -- 14.78s on June 1st.  

The results of the model runs for Bailey are listed in 
Tables~\ref{3db150},\ref{4db150}, and are broken up into 50m splits.   
Since the model 
calculates ``raw'' race times, a reaction time must be added on (I've
assumed a +0.170s reaction, similar to Atlanta).  The ``final'' race
times are in the last column of the tables.

For the best Bailey
guess ($p=0.60$), he takes Johnson regardless of lane choice (assuming
14.73s is an overestimate for Johnson).  However, for larger $p$, which
may be more realistic, Bailey's lane choice starts to become crucial.
That is, for $p=0.70$, he can only win if he's assigned the outside lane.
In the worst case, Bailey gets edged out, drained from fighting the
curve.

\section{How about 200m?}

This model could also be used to predict possible 200m times which 
Bailey might be able to run.  His PBs are recorded as 20.76s, and
20.39s (wind--assisted), both in 1994.  Over the span of 3 years, it's most 
likely the case that his overall endurance has increased, and that he
would be capable of running in the range of 20.30s (again,
bearing in mind that his training is as a 100m specialist).  

Tables~\ref{200outdoors} and ~\ref{200indoors} show predictions for
outdoor and indoor races, respectively.  Assuming the target range described
above, then outdoor $p$ values of 0.50 -- 0.70 provide realistic estimates
(20.28s -- 20.58s).  Meanwhile, for indoor venues (Table~\ref{200indoors}), 
lower values of $p$ give believable times.  These are slower than the outdoor 
predictions, as one might expect, yet still within the grasp of the 100m 
champion (around 20.47s--20.86s).  

The higher values of $p$ for the indoor track
give what are certainly inaccurate times, and hardly world class.  
Recall, though, that the
centrifugal force depends on $1/R$; a sprinter traveling at the same
speed over a radius half as big feels twice the force.  Since $p$ is
the square of the percentage of force felt, then a ratio of 4:1 for 
outdoor:indoor values of $p$ seems reasonable.  This is
why tracks are banked, so the sprinter doesn't have to expend too much
energy to compensate for the curve (unless the race is run in Sherbrooke,
where the sprinter must fight not to fall into the center!).  

\section{And the winner is...}

So, what is the end results of all this?  Will Bailey win the 150m 
showdown, or will Johnson?  It all depends on how each handles the turn,
their lane assignments, and their reaction times (which, in essence, are the
factors that determine the winner in any race!).   Realistically,
people aren't machines that abide by equations,  so the model 
doesn't pretend to 
say how Bailey will {\it definitely} run. What it does
show, though, is that the race is not a clear--cut victory by either party: it 
literally could come right down to the wire.  If each performs at
their Atlanta prime (or better), then I, for one, will be on the
edge of my seat at Skydome for those 14.?? seconds! 

What will the victory signify?  Should
Johnson prevail, would he usurp Bailey's title of World's Fastest
Man?  In my opinion, no.  Bailey won the traditional event to claim
the title, set a world record, and achieved a higher speed than
Johnson.  American sour--grapes aside, this spectacle will only serve 
to show who would win over 150m.  And truthfully, the winners will be 
Bailey {\it and} Johnson, who will walk
from Skydome a combined \$1.5M richer than they were the day before.
Canadian Track and Field will win, because the event will hopefully
regenerate significant interest in the sport.  Finally, the audience
will win, because they will be treated to a magnificent race between
two of history's greatest sprinters.
  
\pagebreak

\begin{table}
\begin{center}
{\begin{tabular}{|l||c c c c c c c c c c|}\hline
Split&10m&20&30&40&50&60&70&80&90&100m \\ \hline\hline
Speed&9.32&10.95&11.67&11.99&12.10&12.10&11.99&11.85&11.67&11.47 \\ \hline
Raw&1.89&2.90&3.79&4.64&5.47&6.29&7.12&7.96&8.81&9.67 \\ \hline
$+$reaction& 2.06&3.07&3.96&4.81&5.64&6.46&7.29&8.13&8.98&9.84 \\ \hline
Official&1.9&3.1&4.1&4.9&5.6&6.5&7.2&8.1&9.0&9.84 \\ \hline
\end{tabular}}
\end{center}
\caption{Predicted splits (s) and speed (m/s) compared with official for 
Bailey's 100m final in Atlanta.  Reaction time is rounded to $+$0.17s.}
\label{100splits}
\end{table}

\begin{table}
\begin{center}
{\begin{tabular}{|c|c|c|c|c|}\hline
$p$&$t_{50}$&$t_{100}$&$t_{150}$&$t_{150}+0.170$\\ \hline
0.60& 5.62 & 9.95 & 14.57 & 14.74 \\ \hline
0.70 & 5.64 & 9.99 & 14.61 & 14.78 \\ \hline
0.80 & 5.67 & 10.03 & 14.66 & 14.83 \\ \hline
\end{tabular}}
\end{center}
\caption{Bailey's predicted Skydome 150m times for various values of $p$,
assuming race is run in lane 3.}
\label{3db150}
\end{table}

\begin{table}
\begin{center}
{\begin{tabular}{|c|c|c|c|c|}\hline
$p$&$t_{50}$&$t_{100}$&$t_{150}$&$t_{150}+0.170$\\ \hline
0.60& 5.61 & 9.93 & 14.54 & 14.71 \\ \hline
0.70 & 5.63 & 9.97 & 14.59 & 14.76 \\ \hline
0.80 & 5.65 & 10.00 & 14.63 & 14.80 \\ \hline
\end{tabular}}
\end{center}
\caption{Bailey's predicted Skydome 150m times for various values of $p$,
assuming race is run in lane 4}.
\label{4db150}
\end{table}

\begin{table}
\begin{center}
{\begin{tabular}{|c|c c|c c| c| c| c|}\hline
$p$&$t_{50}$&$v_{50}$&$t_{100}$&$v_{100}$&$t_{150}$&$t_{200}$&$t_{200}+0
.15$\\
\hline
0.25 & 5.53 & 11.74 & 9.89 & 11.03 & 14.56 & 19.81 & 19.96 \\ \hline
0.36 & 5.55 & 11.60 & 9.98 & 10.85 & 14.69 & 19.96 & 20.11 \\ \hline
0.50 & 5.59 & 11.43 & 10.09 & 10.65 & 14.84 & 20.13 & 20.28 \\ \hline
0.60 & 5.61 & 11.31 & 10.16&10.51&14.93&20.24&20.39 \\ \hline
0.70 & 5.63 & 11.20 & 10.24 & 10.39 & 15.09 & 20.43 & 20.58 \\ \hline
\end{tabular}}
\end{center}
\caption{Bailey's predicted outdoor 200m times, as run in lane 4.}
\label{200outdoors}
\end{table}
 
\begin{table}
\begin{center}
{\begin{tabular}{|c|c|c|c|c|c|}\hline
$p$&$t_{50}$&$t_{100}$&$t_{150}$&$t_{200}$&$t_{200}+0.15$\\ \hline
0.20& 5.62 & 9.91 & 14.88 & 20.32 & 20.47 \\ \hline
0.30& 5.68 & 10.01 & 15.17 & 20.71 & 20.86 \\ \hline
0.40& 5.75 & 10.13 & 15.43 & 21.05 & 21.20 \\ \hline
0.50& 5.81 & 10.22 & 15.67 & 21.37 & 21.52 \\ \hline
0.60& 5.88 & 10.32 & 15.91 & 21.68 & 21.83 \\ \hline
0.70& 5.94 & 10.42 & 16.13 & 21.97 & 22.12 \\ \hline
0.80& 5.99 & 10.50 & 16.33 & 22.23 & 22.38 \\ \hline
\end{tabular}}
\end{center}
\caption{Bailey's predicted indoor 200m times, as run in lane 4.}
\label{200indoors}
\end{table}

\end{document}